\documentclass{sig-alternate}

\usepackage{listings}

\makeatletter
\def\@copyrightspace{\relax}
\makeatother

\begin{document}

\title{Taming the Concurrency:\\Controlling Concurrent Behavior while Testing Multithreaded Software
\titlenote{This research was supported in part by THE ISRAEL SCIENCE FOUNDATION (grant No. 476/11 )}}

\numberofauthors{2} 

\author{
\alignauthor
Evgeny Vainer\\
       \affaddr{The Blavatnik School of Computer Science}\\
       \affaddr{Tel Aviv University}\\
       \affaddr{Tel Aviv, Israel}\\
       \email{zvainer@post.tau.ac.il}
\alignauthor
Amiram Yehudai\\
       \affaddr{The Blavatnik School of Computer Science}\\
       \affaddr{Tel Aviv University}\\
       \affaddr{Tel Aviv, Israel}\\
       \email{amiramy@post.tau.ac.il}
}

\maketitle

\begin{abstract}

Developing multithreaded software is an extremely challenging task, even for experienced programmers. The challenge does not end after the code is written. There are other tasks  associated with a development process that become exceptionally hard in a multithreaded environment. A good example of this is creating unit tests for concurrent data structures. In addition to the desired test logic, such a test contains plenty of synchronization code that makes it hard to understand and maintain.

In our work we propose a novel approach for specifying and executing schedules for multithreaded tests. It allows explicit specification of desired thread scheduling for some unit test and enforces it during the test execution, giving the developer an ability to construct deterministic and repeatable unit tests. This goal is achieved by combining a few basic tools available in every modern runtime/IDE and does not require dedicated runtime environment, new specification language or code under test modifications.

\end{abstract}

\category{D.2.5}{ Testing and Debugging }{Debugging aids}
\category{D.3.3}{ Language Constructs and Features }{Concurrent programming structures}

\terms {Algorithms, Languages}

\keywords {concurrent code, unit test, multithreaded, thread scheduling,
bug reproduction}

\section{Introduction}
\label{ch:Introduction}

In recent years multicore hardware has become a commodity in end user products. In order to support such a change and to guarantee better performance and hardware utilization, more and more application developers had to switch to using multiple threads in their code. Developing such a code introduces new challenges that the developer has to cope with, like multiple threads  synchronization or data races, making concurrent applications much more difficult and complicated to create, even for experienced developers \cite{lu08, long03}. Fortunately, during the years, a lot of tools supporting development process has been created - starting with new synchronization primitives and concurrent data structures and including frameworks that fully isolate all the multithreaded work from the developer.

Another challenge the developer has to face while creating concurrent application is its testing and validation. While testing ``traditional'' single threaded application, the tester is usually able to reproduce the bug by providing the application some constant set of input parameters. This capability allows him, for example, to create a test (or unit test) that demonstrates some buggy behavior and later on use it to validate that the bug was fixed. Unfortunately, such a useful property of the bugs disappear when switching to multithreaded code. In fact, the result of some multithreaded code strongly depends on the context switches that happened during the run, while the developer has almost no ability to control or even predict them \cite{yang99, radnoci09}. This kind of ``non determinism'' during the tests run makes concurrent code very hard to check - some test may always pass on the developer's machine or team's test server but always fail in end user's environment. 

To overcome this problem, unit tests developers try to force context switches in the critical code regions or to delay some code block execution until another code block execution ends. These goals are usually achieved by adding additional operations (like Sleep or Wait/Notify) to the test logic, thus making the test more complicated. This approach creates additional problems. The sleep intervals are usually chosen by trial and error, and there is no guarantee that the next run will pass even if there is no bug. Using Wait/Notify pair instead of Sleep method usually requires modifications in the code under test, since the test scheduling almost always depends on its state (i.e. test code should wait until code under test will enter some state). But even this is often not enough, since in many cases the test failure depends on a context switch that should happen in some third party component. In such a case, the developers have no convenient way to reproduce the bug. 

This problem is well known, and many papers and tools have tried to simplify concurrent code testing \cite{eytani05, souza11}. These papers try to apply very powerful techniques like static and runtime analysis or context switch enumeration in order to decide whether or not some concurrent code is buggy. Although these techniques are very powerful, the problem the authors address is very complex. As a result, none of these works can propose a complete solution. There are many interesting and promising results (we mention some of them in the Related Works section), but more work is required. The authors of these techniques have to overcome such challenging problems like scale, precision rates (both for false positives and false negatives) and extend their methods to the whole set of synchronization primitives existing in modern languages.

In this work, we propose another approach to the given problem. Instead of solving a very general question whether a given code is correct, we want to give the developers an ability to control the thread scheduling during the test run. In other words, if the success or failure of the test depends on the context switches that occur during the test run, then include the desired schedule as part of the test set up. To demonstrate and evaluate our ideas we implemented a framework called \textit{Interleaving} using the Java programming language. Our framework allows the developers: 

\begin{itemize}
	\item to introduce context switches in any arbitrary place in the code, including code under test and third party libraries

	\item to delay some code block execution until some other code reaches the desired state

	\item to reproduce buggy behavior in a deterministic way 
	
	\item to separate all scheduling logic from the test's functional logic
\end{itemize}

These capabilities are achieved by combining together a few simple tools most of the developers are familiar with, so that there is no need for code under test modifications, special runtime or a new language to define the schedule. In addition, our work is based on ideas and tools that exist in every modern platform and IDE and it has no strict dependences on JRE, so a similar framework could be easily implemented for other development platforms.

The rest of the paper is organized as follows:

\begin{itemize}
	\item section \ref{solution} gives more detailed description of our idea, including some implementation details
	
	\item section \ref{implementation} describes our prototype implementation

	\item section \ref{evaluation} provides evaluation of the  \textit{Interleaving} framework

	\item section \ref{relatedworks} reviews some other work in this area

	\item section \ref{conclusions} concludes and provides some ideas for future research
\end{itemize}

\section{Idea}
\label{sec:gatedefinition}
\label{solution}
To achieve such challenging goals we would like to define a new concept we call Gate. For now, it is an abstract concept and its implementation in Java environment will be discussed later in this paper.

\newdef{definition}{Definition}
\begin{definition} 
\label{def:gatedefinition}  Gate  $\mathcal{G} = <\mathcal{L},\mathcal{C}>$ 
where: 

\begin{itemize}

\item $\mathcal{L}$ - some location in code which the execution flow could reach during the
test run

\item $\mathcal{C}$ - some boolean condition that evaluates to true or false

\end{itemize}
\end{definition}

The intuition behind this definition is as following - like any gate
in the real world that has a location it is placed in and could be opened
or closed, our \textit{Interleaving} gate is placed somewhere in the code ($\mathcal{L}$)
and could be opened ($\mathcal{C}$ evaluates to true) or closed ($\mathcal{C}$ evaluates to
false). 

Please note that the latter definition does not limit the position
of a gate in any way. The gate could be placed anywhere - in the code
of the test, in the code under test or even in some third party library.
Furthermore, the gate does not have to be bound to a specific line
of code. Its position could be defined in some other way like ``the
first time method X is invoked'' or ``the fifth iteration of loop P''. 

The same remark holds for condition $\mathcal{C}$ - it could check anything 
one wants. For example, some condition could evaluate to true
only if the time of the day is between 8:00 AM to 5:00 PM while another
one will be true only if it rains outside. Of course, such strange conditions
will have no value for real tests and its more likely that the test
developers will be interested in conditions like ``thread X passed
line Y of the code'' or ``object O is in state S''. 

While executing the test, the execution flow of some thread $\mathcal{T}$ could
reach the location $\mathcal{L}$. At this point the execution of $\mathcal{T}$ is suspended
and condition $\mathcal{C}$ is evaluated. The following behavior of $\mathcal{T}$ depends
on $\mathcal{C}$'s value:
\begin{itemize}
\item $\mathcal{C}$ evaluates to true - thread $\mathcal{T}$ is resumed and continues its execution
in a regular way
\item $\mathcal{C}$ evaluates to false - thread $\mathcal{T}$ remains suspended and will be resumed
only after the value of $\mathcal{C}$ changes to true
\end{itemize}
For now, we are not interested in the mechanism used to notify the
runtime about the changes in condition's state. Let us just assume that
such a mechanism exists and that thread $\mathcal{T}$ will be resumed as soon
as $\mathcal{C}$'s value will change to true. 

Now assume that the unit test developer has an easy and convenient way
to define the gates (both location and condition), to combine them
into sets and to bind these sets to a specified test. Such a powerful
tool will allow the developer to enforce any thread scheduling he
wants. All one needs to do is to identify the code blocks that
should be executed in a particular order and define the gate before
the latter (second) block that will open only after
execution of the first block is completed.

To demonstrate this idea let us assume the example in the Java
programming language shown in figure \ref{ex:sharedmemory}.

\begin{figure}
\centering

\begin{lstlisting}
public class SharedMemoryAccessExample { 
	int multiplier = -1; 	
	int result = 0;

	class Worker1 extends Thread { 
		public void run() { 
			multiplier = 1; 
		} 
	} 

	class Worker2 extends Thread {
		public void run() {
			result = multiplier * 10; 		
		}
	}
	
	public int Calculate() throws Exception { 
		Thread t1 = new Worker1(); 
		Thread t2 = new Worker2();

		t1.start(); 
		t2.start();
		t1.join();
		t2.join();
		return result;
	} 
}

\end{lstlisting}

\caption{Shared Memory Access}
\label{ex:sharedmemory}

\end{figure}

In this very simple example each call to the Calculate method will cause
the runtime to create two threads, execute them and return the value
stored in the variable ``result''. One could easily note that the
value returned by Calculate method depends on the order in which the worker
threads were executed. Let us assume that the expected result is 10, while the result -10 (which will be returned if line 13 executed
before line 07) is a bug. 

Even such a simple example of a multithreaded class could be very difficult
to test. Following the encapsulation principle of OOP all the members
of this class are internal, so the unit test code that is external to the class 
has no access to them. As a result, the only thing the unit test developer
could do is to call the Calculate method and to check its return value.
It is obvious that the outcome of such a test will depend on the thread
schedule that took place during the test run. Such a unit test has
no value at all since its outcome is not deterministic and the fact that
the test passed does not guarantee that the code is bug free.
One could try to increase the confidence of the test by calling the Calculate
method multiple times during the test and validating all the values
returned. Such a test will not be much better than the previous version
since it still can result in false negative.

Now assume that the unit test developer is able to define gates as
described before. In such a case, one could define the
gate
\begin{displaymath}
\mathcal{G} = <line\ 07,\ thread\ Worker2\ finished\ its\ execution>
\end{displaymath}
and bind
it to the test. According to the semantics of the gates defined earlier,
doing so will cause Worker1 thread to pause its execution just before
line 07 of the code and to remain suspended until Worker2 is done.
As a result, a call to the Calculate method will return -10, thus failing
the test. This thread ordering will be constantly enforced every
time the test will be executed, allowing the developer to reproduce
the buggy behavior in a deterministic way.

\section{Implementation}
\label{implementation}
In order to demonstrate and evaluate our ideas we implemented the
above concept using the Java programming language and JRE environment.
The resulting framework, called \textit{Interleaving}, provides an
ability to place the gates in arbitrary places in code and to evaluate
the conditions when the gate is reached, forcing the behavior defined
earlier. The framework could be used together with Eclipse IDE, providing
the developers familiar and convenient environment to define and manage
their gates. Of course, the concept of a gate defined earlier is very
general, so we had to make some relaxations while implementing it.

\subsection{Condition definition}

First of all, in our implementation, we decided to utilize Java programming
language for gate conditions definitions. There are several advantages
for such a choice:
\begin{itemize}
\item Java is a very powerful programming language. Any
special language we could create for condition definitions would be
less expressive than Java.

\item JRE contains a lot of frameworks and code libraries. All of them could be used while defining gate conditions.
This simplifies conditions' definitions and allows the developers
to create more complicated gates.

\item Using language the developers are familiar with to define gate conditions significantly simplifies
migration to our framework.

\item Using Java for conditions definitions allows us to use JRE in order
to evaluate condition's value.

\item The fact that conditions are defined using the same programming language
that was used while developing the application makes the conditions
much more powerful. For example, the code in gate condition can
interact with objects defined in the application, check their states or
even call their methods. All of this is possible because the same
language is used to define conditions and application and because the same runtime is used to execute them.
\end{itemize}

Using Java for conditions definitions limits
the power of gates, with respect to the definition given in section \ref{sec:gatedefinition}.
Nevertheless, the code under test is created using the same programming
language and executed using the same runtime engine as \textit{Interleaving}'s
gates' conditions. This observation refines the fact that the gates 
at least as powerful as the application itself, 
justifying this implementation decision.

\subsection{Notification mechanism }

\label{sec:notificationmechanism}

Another implementation decision we made deals with the gate notification
mechanism. As section \ref{sec:gatedefinition} states, if some
thread $\mathcal{T}$ is suspended on gate $\mathcal{G} = < \mathcal{L}, \mathcal{C} >$, it is resumed immediately when
$\mathcal{C}$'s value becomes true. This definition assumes some mechanism that
observes the value of the condition all the time and is able to resume $\mathcal{T}$ whenever
condition state changes. Although it is possible to implement
such a mechanism, the implementation may be pretty complex and
somewhat tricky. Since the purpose of our implementation is to demonstrate the
ideas and not to provide market ready solution, we decided to simplify
this behavior. In the \textit{Interleaving} framework, the implementation of
the notification mechanism is part of the condition's logic and is the responsibility of the test developer.
In other words, when thread $\mathcal{T}$ reaches gate $\mathcal{G} = < \mathcal{L} , \mathcal{C} >$ its state $\mathcal{S}$
is saved somewhere aside and the condition's logic is evaluated. This
evaluation should return only after the gate is considered to be opened.
After the condition's evaluation ends, the thread's state $\mathcal{S}$ is restored
and $\mathcal{T}$ continues its execution in the regular way. This behavior fits the
gate's behavior from section \ref{sec:gatedefinition}, since thread
$\mathcal{T}$ can not continue it's execution until $\mathcal{C}$ is satisfied. Since conditions'
logic is defined using Java programming language, it is not a problem
to create such complex conditions.

This relaxation allows test developers to define different and complex
conditions whose behavior depends on test requirements. From the observations
we made while evaluating our framework, most of the test scheduling
could be created using very simple ``manual'' gates, i.e. the gates
whose state has to be changed explicitly. The condition of such a
gate contains one expression only - calling for Wait method on some
object, while appropriate Notify call has to be made explicitly somewhere
else in the code. Please pay attention that such a call could be placed
anywhere in the code (even in third party libraries) using fictitious
gate whose condition contains Notify call only. Of course, as we mentioned
earlier, more complex conditions could be introduced in order to create
more complex schedules. Some examples of such conditions will be discussed 
in section \ref{evaluation}.

\subsection{Location definition}

Now we describe the technique we used to define the location $\mathcal{L}$ for
some gate. While developing \textit{Interleaving} framework we searched
for a way to represent the location that will satisfy the following
requirements:
\begin{itemize}
\item The test developer should have fine grained control over gates positions,
i.e.  one should be able to bind the gate to some line in the source
code, to some instruction in the binary file or, if possible, to some
event that happens during the application execution (like first
exception thrown or entering some method).

\item The framework should be able to intercept the execution flow of any
thread that reaches the location defined by some gate $\mathcal{G}$ in order to
evaluate the condition and suspend thread's execution if needed.
\end{itemize}

Fortunately, we are not the first who looked for such capabilities.
The entity that satisfies these requirements was invented long ago
and already exists in all modern development languages and
platforms - it is a breakpoint. Indeed, the breakpoint mechanism of
JRE allows the developer to put the breakpoint in almost arbitrary
place in the code, including third party libraries. It also supports
more complex conditions like hits counter, method entry/exit or class
load events. Every modern IDE (like Eclipse, for example) provides
the developer some convenient, usually graphic, interface for breakpoint
definition, fully abstracting from the real syntax used to define
breakpoint location/condition. On the other hand, Java Debugging Interface
(JDI) libraries supported by the last versions of JVM provide very
powerful programmatic interface which allows us to define and remove
breakpoints, receive notifications when some breakpoint is hit and
execute some custom action when this happens. All of this makes a
breakpoint mechanism an ideal solution for defining gates' locations.

\subsection{Flow control}

\label{sec:flowcontrol}

We now present a short description of the technique the
\textit{Interleaving} framework uses in order to intercept and control
the flow of test execution. 

Each \textit{Interleaving} test is a simple JUnit test while we use
JUnit rules to enrich its functionality. At runtime, JUnit will discover
that the test has additional rule and will pass the control to this
rule. This is how \textit{Interleaving} comes into the game. The rule
code will investigate current test and locate the gates relevant for
the test (the way we associate gates to tests is described later
in section \ref{sec:userinterface}). Next, a few things will happen. 
\begin{itemize}
\item First, \textit{Interleaving} will compile the Java code defined
in gates' conditions fields, creating a separate static method for each
one of the gates. 
\item Next, \textit{Interleaving} uses JDI to set the breakpoints in all of
the code locations defined by the gates, and starts a special thread
that will handle those breakpoints hits. 
\end{itemize}
After this work is done, the rule returns the flow to JUnit and it
continues test execution in a regular way. 

While running the test, some of the breakpoints might be hit. When
this happens, the thread $\mathcal{T}$ that hit the breakpoint is suspended by the 
JVM (all other application threads continue to run) and a notification
is sent to the special \textit{Interleaving} thread mentioned earlier. The
notification contains all the necessary information required by \textit{Interleaving}
in order to identify the gate that was reached and to locate a
method containing the gate's condition's code. Next, this method is placed
on top of $\mathcal{T}$'s stack and $\mathcal{T}$ is resumed. This technique causes $\mathcal{T}$
to leave the state it was in when it hit the breakpoint, and forces
it to execute new code - the code of the condition the developer supplied.
Moreover, when the condition's code will return, the stack frame of the 
condition's method will be destroyed and the thread will return to
the same state it was in when it was suspended. Since the thread is
not suspended anymore it continues the execution of the original test
logic as if nothing happened. The only side effect one could
notice is a delay caused by the condition's evaluation. This delay, combined
with the condition's behavior defined earlier (section \ref{sec:notificationmechanism}),
gives us all we need to enforce the desired scheduling. 

It is important to notice that all the operations described in the
current section are achieved using standard APIs and extension points
provided by JUnit, JVM and JDI library. At the cost of some additional
code written, we manged to implement these capabilities without modifications
made to any of those libraries. As a result, the \textit{Interleaving} framework
does not require special versions of JVM or JRE in order to run the
tests. The tests can be executed using the same environment that
is used in the production stage.

\subsection{Putting everything together}
\label{sec:alltogether}

Now, we would like to describe how all the things we mentioned earlier
are combined together in the \textit{Interleaving} framework. For the
demonstration purpose, we assume some developer is required to create
a test that reproduces a concurrent bug that exists in the code of figure \ref{ex:sharedmemory}.
After investigating the bug, the developer concludes that the bug
happens only if line 13 of code is executed before line 07, so while
creating the test he needs to enforce this schedule. 

To do so, he will have to use one of the gates defined in \textit{Interleaving}
framework named ``SimpleGate''. This gate defines a simple API composed
of two methods - Wait and Open. Each SimpleGate instance maintains
some internal condition that initially evaluates to false (i.e. the
gate is considered to be closed) and it remains so until the Open
method is called. Calling this method changes the internal condition's
value in such a way that from this point it always evaluates to true
(i.e. the gate is considered to be open) and there is no way to switch
the gate back to the closed state. The Wait method of the gate implements
the notification logic we described earlier in section \ref{sec:notificationmechanism}.
Whenever this method is called, it returns only after the gate's instance
it was called on is in opened state. Using this gate the developer
can ensure that the code block following the gate's Wait call will
be executed only after the code block preceding the gate's Open call
is done. 

Now, in the test, the developer has to create an instance of SimpleGate
and give it some meaningful name, ``Worker2\ Done'' for example.
Next, he has to locate it somewhere in the code. Following the example,
he wants to suspend the execution of the code on line 07 so this is
the line where the gate should be located. In order to mark this line
as a gate location the developer puts a breakpoint on it. Now,
he has to specify the condition associated with the breakpoint. For
this purpose we decided to utilize the conditional breakpoint window
of Eclipse IDE. So, the developer marks the earlier created breakpoint
as conditional one and in the condition window writes the code that calls
for Wait method of ``Worker2Done'' gate. Next, he has to choose
the point where the gate is to be opened. Obviously, this point is at line 14
(alternatively, it might be the point where some thread finishes
the execution of Worker2.run method). So, the developer puts another
breakpoint on line 14 (or method exit breakpoint on Worker2.run method),
marks it as conditional and writes the condition that calls for ``Worker2Done''
gate's Open method. The combination of these two breakpoints creates a
deterministic schedule which always enforces the code at line 07 to
run after the code at line 13. 

Now, all that is left is to write the test that calls for Calculate method
and to associate the gates created earlier to this specific test. This
association could be done using Working Sets. Working set is a convenient
way the Eclipse IDE provides for the purpose of grouping some related
entities of any kind. All the developer has to do in order to associate the
gates with the test is to create breakpoints working set, give it a name
of the test and add the  breakpoints created earlier to this set. Now,
the test can be run using standard JUnit test runner.

While executing the test the breakpoint set on line 07 will be hit
by thread $\mathcal{T}_1$. At this point, \textit{Interleaving} will use the technique
we described in section \ref{sec:flowcontrol} to cause $\mathcal{T}_1$ to execute
the breakpoint's condition. This condition contains the call to Wait
method of ``Worker2Done'' gate. As we recall, the Wait
method of the gate will return only after the Open method of the same
gate was called. Let us assume that the Open method of ``Worker2Done'' gate was not called yet.
Therefore, $\mathcal{T}_1$ will remain inside the code of Wait method, while all
the other application threads will execute the test logic in the regular
way. At some point of time, some other thread $\mathcal{T}_2$ will hit the breakpoint
located at line 14, this will cause $\mathcal{T}_2$ to stop its current flow execution
and to execute the code defined by the condition of this breakpoint and,
as a part of it, to execute the call for Open method of ``Worker2Done''
gate. This call will return immediately allowing $\mathcal{T}_2$ to return to the test
logic. In addition, this call will cause the Wait method of ``Worker2Done''
gate to return, releasing $\mathcal{T}_1$ and allowing it to return to the test logic
execution. 

As a conclusion of the flow described, one can notice that adding
gates to the test introduced some new ordering constraints on events
that occur during the test run. These constraints are as follows 
(we use the notation of $\mathcal{E}_1 \rightarrow \mathcal{E}_2$
to denote that event $\mathcal{E}_1$ occurs before event $\mathcal{E}_2$):

\begin{itemize}
\item The code in line 13 is executed ($\mathcal{A}$) before the breakpoint on line 14 is hit ($\mathcal{B}$)
($\mathcal{A} \rightarrow \mathcal{B}$)
\item ``Worker2Done'' Open method is called ($\mathcal{C}$) after the breakpoint on line 14 is hit 
($\mathcal{B} \rightarrow \mathcal{C}$)
\item ``Worker2Done'' Wait method returns ($\mathcal{D}$) after its Open method is called
($\mathcal{C} \rightarrow \mathcal{D}$)
\item condition evaluation in $\mathcal{T}_1$ ends ($\mathcal{E}$) after ``Worker2Done'' Wait method returns 
($\mathcal{D} \rightarrow \mathcal{E}$)
\item thread $\mathcal{T}_1$ returns to test logic execution ($\mathcal{F}$) after it completed condition evaluation 
($\mathcal{E} \rightarrow \mathcal{F}$)
\item the breakpoint in line 07 is hit before the code on the same line
is executed, as a result $\mathcal{T}_1$ will execute the code in line 07 ($\mathcal{G}$) only
after it returns back to the test logic evaluation 
($\mathcal{F} \rightarrow \mathcal{G}$)
\end{itemize}

Events sequence above implies that $\mathcal{A} \rightarrow \mathcal{G}$ (i.e. the code in line 13 will always be executed before 
the code in line 07), resulting in consistent bug reproduction, no matter 
what was the threads scheduling created by JVM/OS for the current test
execution. 

\subsection{Deadlock detection}

\label{sec:deadlockdetection}

As one could already notice, using \textit{Interleaving} framework means interfering with  threads scheduling. This is what the framework was created for and this is where its additional value comes from. However, threads synchronization is a very delicate area. Careless positioning of the gates inside the code or incorrect use of notification mechanisms may lead to deadlocks that otherwise would never arise in the original code. 

In order to cope with this problem, every framework like \textit{Interleaving} has to provide some deadlock detection mechanism that will break the test execution and notify the tester as soon as the deadlock discovered.  The logic of such a mechanism is the separate topic many researches address \cite{stoller, pulse} and, in our opinion, is out of the scope of our research. 

However, we believe that any production ready tool should incorporate known techniques for deadlock detection to complement its ability to control the schedules.


\subsection{User interface}

\label{sec:userinterface}

One of the things we always kept in mind while creating the \textit{Interleaving}
framework is its usability. Providing the developers with a tool that is
based on concepts they are familiar with significantly reduces the
learning curve and eases the migration. Till now we described two
examples of such a reuse in our framework:
\begin{itemize}
\item using Java programming language in order to describe gates' conditions
\item using breakpoint mechanism in order to define gates' locations
\end{itemize}

Another example of this approach is the user interface of the \textit{Interleaving}
framework. All the operations the test developer has to perform while
creating and executing interleaved test could be done using standard
Eclipse IDE environment and no additional plugins/windows are required.
In our opinion such an integration is very important, since the developer
fills comfortable with the environment and can focus on his actual
job, instead of spending time on learning new concepts.

\section{Evaluation}

\label{evaluation}

The evaluation of our work consists of two parts. First, we looked
for different examples of concurrent bugs that are hard to reproduce
using standard testing tools and created the gates sets that reproduce
the buggy behavior in a consistent way. A few such examples are presented
in this section. Some of them are real bugs taken from the bugs repositories,
while others are synthetic examples we created in order to demonstrate
the expressiveness and the power of our approach. The second part of the
evaluation is done via the comparison to other works. We show that
our framework is at least as powerful
as some other tools presented in recent papers, 
and in some cases more powerful.

\subsection{Examples}

\subsubsection{Unspecified Time}

\label{sec:time}

This example is a synthetic one, but it demonstrates a very common
scenario. Suppose the tester needs to check a class that performs
some long time operation in a different thread. The amount of time the
operation could take varies from run to run in hardly predictable
way, and depends mostly on the environment the test is run on. In order
to create such a test, the developer needs to execute the operation,
wait until the job is finished and only then check its status. Figure
\ref{ex:longrunningjunit} contains sample code that demonstrates this approach. 

\begin{figure}
\centering

\begin{lstlisting}
@Test
public void LongRunningTask_JUnit()
						 throws Exception {
	Task task = new LongRunningTask().new Task();
	
	task.start();
	
	Thread.sleep(task.MaxTime);
	assertTrue(task.IsDone);
}
\end{lstlisting}
\caption{Unit test for LongRunningTask class}
\label{ex:longrunningjunit}
\end{figure}

In this example, we assume that the operation time is upper bounded by some
constant. If it is not the case, the test could ``busy wait'' until
the operation is done. Both methods are not perfect - in the former case
the test always takes the maximal possible time even when the operation
ends very fast, while the ``busy wait'' option consumes unnecessary
machine resources.

\begin{figure}
\centering
\begin{lstlisting}
@Test
@Interleaved
public void LongRunningTask_Interleaved() 
						 throws Exception {
	Task task = new LongRunningTask().new Task();
	
	task.start();
	
	//interleavings.GateManager.Wait("task_done");
	assertTrue(task.IsDone);
}

\end{lstlisting}

\begin{lstlisting}[mathescape, numbers=none, breakindent=20pt, xleftmargin=0pt ]
$\mathcal{G}_1 = <$ LongRunningTask_Interleaved@10, interleavings.GateManager.Wait("task_done"); $>$ 
$\mathcal{G}_2 = <$ LongRunningTask@33, interleavings.GateManager.Open("task_done"); $>$ 
\end{lstlisting}

\caption{Unit test and gate for LongRunningTask class using Interleaving framework}
\label{ex:longrunninginterleaved}
\end{figure}

Figure \ref{ex:longrunninginterleaved} demonstrates \textit{Interleaving} version of such a test. It contains two gates:
\begin{itemize}
\item $\mathcal{G}_1$ is located just before the assertTrue call. The gate remains closed
until the task is done (optionally this gate could be removed from
the test set up and replaced by the commented line)
\item $\mathcal{G}_2$ is a fictitious gate (as described in section \ref{sec:notificationmechanism})
that opens $\mathcal{G}_1$ and is located on the last line of the checked operation
(line 33 of LongRunningTask.java  \footnote{the source code is not listed in the paper})
\end{itemize}
Using this technique the test gets the best of the two worlds -- it
takes as little time as the checked job takes, and the test thread is
blocked while the operation performs. In addition, in case the operation
class would not provide us with MaxTime and IsDone members, the developer
has no convenient way to check this scenario without using \textit{Interleaving}
capabilities.

\subsubsection{StringBuffer}

\label{sec:stringbuffer}

Our next example deals with a real bug that exists in StringBuffer
class in the current version of JRE \cite{javabug1, javabug2, park}.
Figure \ref{ex:stringbuilder}  contains the code of the append method of AbstractStringBuilder
class which StringBuffer class inherits. 

\begin{figure}
\centering
\begin{lstlisting}
AbstractStringBuilder append(StringBuffer sb) {
	if (sb == null)
		return append("null");

	int len = sb.length();
	int newCount = count + len;
	if (newCount > value.length)
 		expandCapacity(newCount);

	sb.getChars(0, len, value, count);
	count = newCount; 
	return this;    
}
\end{lstlisting}
\caption{AbstractStringBuilder.append method}
\label{ex:stringbuilder}
\end{figure}

\begin{figure}
\centering
\begin{lstlisting}
@Test 	
public void Length_Test() throws Exception {
	final StringBuffer sb1			= 
			new StringBuffer("original data");
	final StringBuffer sb2 			= 
			new StringBuffer("appended data"); 			
	
	Thread worker = new Thread(new Runnable() { 
		public void run() {
				sb1.append(sb2);
		} 	
	}); 	
		
	worker.start();
	sb2.setLength(3);
	worker.join();
} 
\end{lstlisting}

\caption{Test method for StringBuffer.append}
\label{ex:stringbuildertest}
\end{figure}

\begin{figure}
\centering
\begin{lstlisting}[mathescape, breakautoindent=true, numbers=none, breakindent=20pt, xleftmargin=0pt   ]
$\mathcal{G}_1 = <$test@15, interleavings.GateManager.Wait("afterget");$>$ 
$\mathcal{G}_{1fictitious}  = <$append@06, interleavings.GateManager.Open("afterget");$>$
$\mathcal{G}_2 = <$append@10, interleavings.GateManager.Wait("afterset");$>$
$\mathcal{G}_{2fictitious} = <$test@16, interleavings.GateManager.Open("afterset");$>$
\end{lstlisting}
\caption{Gates defined for LengthRaceCondition test}
\label{ex:gates}
\end{figure}

This method contains a potential data race while working with the length
of the received argument. If the length of sb changes after line 05 was
performed, but before line  10 is executed, the method could end up
with an exception. One can easily write the test that tries
to reproduce this scenario. An example of such a test is shown in
the figure \ref{ex:stringbuildertest}.

Unfortunately, running this test as is will not reproduce the bug.
The reason for this is that the context switch between the worker thread
and the test thread should happen in a very specific and very short time
window - after the worker thread performed line 05 of append method
but before it reaches line 10 of it. This timing window is pretty
tight and it is very unlikely for the context switch to happen there
in regular runs. The sleeps technique used in many concurrent tests also fails to reproduce the bug.
Usage of this technique requires one of the sleep calls to be located
inside the append method, causing code under test modification which
is undesirable in most cases. In order to reproduce the bug
we tried to execute this test in some different setups - we executed
the test many times inside the loop, we executed several instances
of the test simultaneously, we ran it on different machines under different
loads - all with no success. The bug appeared in very few runs in
a very inconsistent way. The inability to reproduce the bug was noticed
by java developers too. The appropriate bug reports mention that the
bug ``can be reproduced rarely'' \cite{javabug1} and proposes a test containing two
infinite loops (one loop for each thread) \cite{javabug2} in order to reproduce it.

Using \textit{Interleaving} framework we reproduced the buggy behavior
in all of the runs by adding only two gates to the test and without
changing the code at all. The first gate is located in line 15 of
the test and opens after the worker thread passed line 05 of the append
method, while the second is located in line 10 of the append method
and opens after the test performed line 15 of its code. The formal
gates definition is presented in the figure \ref{ex:gates}. 

Please recall that in our implementation all the gates are manual,
i.e. every conceptual gate consists of two parts - the real gate and
some fictitious gate that is responsible for opening the real one, as
described in section \ref{sec:notificationmechanism}

\subsubsection{ArrayList concurrency}

\label{sec:arraylist}

Till now all the examples we presented used SimpleGate in order to define the desired concurrent behavior. Even such a simple gate was powerful enough so that we could reproduce some concurrent bugs that are hard to reproduce using other techniques  existing today. In this example we want to demonstrate the usage of another gate implementing more complicated scheduling logic. 

ArrayList is a well known and widely used class existing in Java. Its current implementation is known to be not thread safe and provide no guarantees 
when using the same ArrayList object in multithreaded environment. Despite this fact, in some cases, the concurrent operations performed on the same 
instance of ArrayList do perform in the expected way since the chance for the data race to happen during the execution is pretty low. This issue may 
confuse inexperienced developers and lead to bugs in the code they write. 

Suppose somebody wants to demonstrate the unsafety of ArrayList in multithreaded scenarios. In order to do this, he needs to perform several operations 
on the same instance of ArrayList class using different threads and ensure that those operations will necessarily lead to object's state corruption. As always, 
doing so is not easy since the context switches that take place during the run are out of control of the test developer.

Figure \ref{ex:addall} presents the code of ArrayList.addAll method. 

As it was mentioned before, this method is not thread safe. For example, if two threads execute addAll code and both of them first perform line 06 of the method 
and only then, simultaneously, execute line 08, the object's state will be corrupted. In order to reproduce this problem one could write a test that is similar to the 
test presented on figure \ref{ex:addalltest}.

The outcome of this test depends on the context switches that took place during its execution. When working on this example we rerun this test many times inside the loop and noticed that the 
assertion on line 31 fails for one execution of several hundreds. 

\textit{Interleaving} allows us to reproduce the race for all test executions. To do so, we created a new gate  we called BarrierGate{\footnote{specific gate implementations are not part of the framework and could be created by testers ``on demand'' according to their needs}}. As follows from its name, the BarrierGate behavior is very similar to the functionality of the synchronization primitive called barrier. In contradiction to the SimpleGate we used earlier, the BarrierGate cannot be opened by calling its Open() method but will open itself after a predefined number of threads (passed as parameter at gate's construction) called its Wait() method. For the test case above, we placed the BarrierGate with the limit of 2 threads on the line 08 of addAll method (figure \ref{ex:addallgates}).

When running the test, the first worker thread reaching line 08 of addAll method will be blocked by the gate until the second worker thread will reach the gate too. At this point of time both worker threads performed arraycopy call (line 06 of addAll method) but none of them increased the internal size variable (line 08 of addAll method), thus the tested object already contains less elements then expected.  After both threads reach the gate, it will open, releasing the workers to perform the rest of addAll method code. Each one of them will increase the size variable, together causing the corruption in the internal state of the ArrayList object they are working on. The assert inside the test code (line 31) will validate the state and fail because of the corruption created by worker threads. This flow will happen for every test execution consistently reproducing the desired bug.

\subsection{Comparison to IMUnit}

\label{sec:imunitcompare}

IMUnit \cite{imunit} is another framework that provides test developers
the ability to define the ordering of some events during test
execution. The scheduling definition for this framework consists of
two parts:
\begin{enumerate}
\item initiation of events of interest somewhere inside the code
\item declarative definition of desired events ordering for the test using
some special syntax
\end{enumerate}
The framework controls tests execution and ensures the desired ordering
in the following manner -- while executing the test, the flow could
reach some event of interest (1) defined by the test developer. At this
moment, the execution of the thread is suspended until all of the
preceding events defined for the test (2) occurred. In addition to
the framework, the authors provide a tool that allows relatively
easy migration from the ``sleep based'' tests to IMUnit notation.
Using this tool the authors succeed to convert a large amount of concurrent
tests to be used with IMUnit, a result that implies the good expressive
power of IMUnit notation.

\begin{figure}
\centering
\begin{lstlisting}
public boolean addAll(Collection<? extends E> c) 
{
	Object[] a = c.toArray();
	int numNew = a.length;
	ensureCapacity(size + numNew);  
	System.arraycopy(a, 0, elementData, size, numNew);
	
	size += numNew;
	return numNew != 0;
}
\end{lstlisting}
\caption{ArrayList.addAll method}
\label{ex:addall}
\end{figure}

\begin{figure}
\centering
\begin{lstlisting}
@Test
public void ArrayList_RaceCondition_Interleaved() 
						 throws Exception {
	final ArrayList<String> tested = 
				new ArrayList<String>();

	Thread worker1 = new Thread(new Runnable() {
		public void run() {
			ArrayList<String> data = 
													new ArrayList<String>();
			data.add("data");
			tested.addAll(data);		
		}
	});
		
	Thread worker2 = new Thread(new Runnable() {
		public void run() {
			ArrayList<String> data = 
													new ArrayList<String>();
			data.add("data");
			tested.addAll(data);		
		}
	});
	
	worker1.start();
	worker2.start();
		
	worker1.join();
	worker2.join();

	assertNotNull(tested.get(tested.size()-1));
}
\end{lstlisting}
\caption{Test for ArrayList.addAll method}
\label{ex:addalltest}
\end{figure}

\begin{figure}
\centering
\begin{lstlisting}[mathescape, breakautoindent=true, numbers=none, breakindent=20pt, xleftmargin=0pt   ]
$\mathcal{G} = <$addAll@08, interleavings.GateManager.Wait("after_copy");$>$ 
\end{lstlisting}
\caption{Gates defined for \newline ArrayList\_RaceCondition\_Interleaved test}
\label{ex:addallgates}
\end{figure}

We will show that IMUnit events are a special case of \textit{Interleaving}
gates and every IMUnit test could be easily rewritten for our framework.
One can immediately conclude that:
\begin{enumerate}
\item the same approach described in \cite{imunit} can be used to convert
the tests to our notation.
\item the expressive power of \textit{Interleaving} notation is at least as
good as that of the IMUnit notation.
\end{enumerate}
Moreover, we will show that the StringBuffer bug mentioned earlier (section \ref{sec:stringbuffer})
can not be reproduced using IMUnit but can easily be reproduced using
\textit{Interleaving}, which implies the greater expressiveness of the \textit{Interleaving}
framework.

In order to substantiate the claims above, we developed a simple algorithm that
allows to convert every IMUnit test to \textit{Interleaving} notation. This algorithm is 
presented in figure \ref{ex:transformation}. We also provide a formal proof that 
the transformation this algorithm applies to the test code does not affect the test result and preserves the scheduling enforced by the framework. Due to space limitations
we will not present this proof here, but only describe the intuition
and the general idea. The whole and formal proof is provided
in \cite{vainer13}. 

\renewcommand\labelenumii{\theenumii }
\renewcommand{\theenumii}{\theenumi.\arabic{enumii}}

\begin{figure}[h]

\begin{enumerate}

\item \textit{let $e_p \rightarrow e_s$ be the IMUnit scheduling defined for the test (which means that event $e_p$ should happen before event $e_s$)}

\item \textit{let $L_{e_p}$ and $L_{e_s}$ be the lines of code where events $e_p$ and $e_s$ are initiated, respectively}

\item \textit{define gate $\mathcal{G}_{e_p \rightarrow e_s} = < \mathcal{L} , \mathcal{C} >$ as follows:}

\begin{enumerate}

\item \textit{$\mathcal{L} = L_{e_s}$}
\item \textit{$\mathcal{C} = L_{e_p}$ $was$ $already$ $executed$}

\end{enumerate}

\end{enumerate}
\caption{Transformation {$\mathcal{T} : IMUnit$ $Tests \rightarrow Interleaving$ $Tests$}}
\label{ex:transformation}

\end{figure}

The intuition behind this transformation is very simple -- the execution
flow could not reach $L_{e_s}$ before it passes the gate $\mathcal{G}_{e_p \rightarrow e_s}$, but the gate
remains closed until the flow executes $L_{e_p}$. This implies that $L_{e_p}$ will
always be executed before $L_{e_s}$, enforcing the desired scheduling. In
the full proof we also show how to transform other types of scheduling
(like ${[}e_p{]} \rightarrow e_s$) and how to deal with complex scheduling that
contains multiple simple scheduling.

Using this simple algorithm one can easily understand why all
the tests created with IMUnit notation are a subset of all the
tests that could be created using \textit{Interleaving}. The reason for that
is that while using IMUnit the events can be initiated from the test
code only, which implies that appropriate gates in the transformed
test will also be placed inside the code of the test (while \textit{Interleaving}
mechanism that uses breakpoints allows the developer to put the gate
almost everywhere - inside the code under test or even in third
parties code). This limitation significantly reduces the set of bugs
IMUnit is capable to reproduce. For example, the StringBuffer bug
mentioned above (section \ref{sec:stringbuffer}) can not be reproduced
using IMUnit because of this issue.

Another conclusion that is immediate from the algorithm above is that
every IMUnit event could be represented using a gate with very simple
and constant condition. This fact also limits the expressive power
of the framework. In order to overcome this limitation IMUnit defines
its own scheduling specification language that allows the developer
to specify more complex condition like ${[}e_p{]} \rightarrow e_s$. The problem
with this approach is that every new condition complicates this language
specification and that test developers have to be familiar with this
language and all of its capabilities. \textit{Interleaving}, in contrast,
does not limit the tester to a predefined set of conditions but
allows him to define every logic he desires using the power of the Java
programming language - the language the developer is already familiar
with. For example, a condition code can check the internal state of
current ``this'' object or even the values of local variables on
the stack, things that are impossible while using IMUnit notation.

\subsection{More Tests}
\label{moretests}
Inspired by the observations presented in section \ref{sec:imunitcompare} we tried to apply them in practice. The IMUnit package available for download at \cite{imunithome} comes with 202 example unit tests that were created by the framework authors based on the real tests from different projects \cite{apache1}, \cite{apache2}, \cite{apache3}, \cite{apache4}, \cite{codehaus}, \cite{jboss}. We converted these tests to the \emph{Interleaving} gates notation by applying a transformation algorithm very similar to one presented on figure \ref{ex:transformation}. The conversion took very little amount of time and effort and at the end we got 196\footnote{there are 6 more tests we did not succeed to execute even in the original IMUnit notation} working interleaving tests that demonstrate consistent behavior for all of the runs. In addition, the outcome of all of the converted test is equal to the outcome of the original tests. Since the origin of all the tests are different real life projects, we conclude that our notation, combined with the prototype implementation we provide, are powerful enough to be used in real life testing. 

\subsection{Runtime Performance}
The performance of unit testing framework is a very important issue. Since many real life projects have thousands of tests, the little overhead the framework creates for each one of the tests can create a huge delay when executing the whole test set. Thus, unit testing framework developers should aim at the lowest overhead they can achieve.

Despite of the statement in the previous paragraph, the first and the most important problem we cope with in this research is the outcome reproducibility and the control we want to give the test developer over the test execution. The prototype implementation we provide with this work was created in order to demonstrate \textit{Interleaving} idea, its feasibility and usability and not in order to compete with mature, production ready frameworks existing today. 

The running time of the tests we measured while using \textit{Interleaving} framework is not significantly different from the running time of the regular concurrent tests validating the same scenario, and in some cases even lower (please recall example \ref{sec:time}). We can report that the average increase in the test execution time is by a factor of x1.05 when using  \textit{Interleaving} framework compared to original IMUnit tests. Table \ref{table:compare} summarizes observed execution time for both frameworks. These results was measured on the test base mentioned in section \ref{moretests}. As reported in \cite{imunit}, switching to IMUnit notation reduces the execution time of the tests by factor of x3.39 when compared to original unit test. Combing these measurements together we can conclude that the execution time of \textit{Interleaving} version of tests is at least by factor of x3 better than this of the original sleep based tests.


\begin{table}[t]
\begin{center}
  \begin{tabular}{ l | r | r || r  }
	
\hline

    Subject & IMUnit \newline [s] & Interleaving \newline [s] & Overhead \\
    \hline \hline

Collections	& 0.114 & 0.124 & 1.08 \\
JBoss-Cache	& 6.097 & 6.094 & 1.00 \\
Lucene 	& 5.641 & 5.906 & 1.05 \\
Pool		& 1.020 & 1.028 & 1.01 \\
Sysunit 	& 0.118 & 0.128 & 1.08 \\
JSR-166 TCK & 2.442 & 2.560 & 1.05 \\

    \hline     \hline

\multicolumn{3}{l ||} {Geometric Mean }   &   1.05 \\

  \end{tabular}
\end{center}
\caption{Test execution time}
\label{table:compare}

\end{table}

\section{Related Works}
\label{relatedworks}

The problem of concurrent software testing has been studied by many researchers and there are plenty of papers and tools addressing different aspects of the problem. Generally, all the  works on the topic could be divided to several groups, according to the approach the authors propose in order to cope with concurrency related issues:

\begin{enumerate}

\item Recording and replaying the error prone run for the future research and debugging

\item Automatic testing of given codebase for the problems caused by concurrency and synchronization issues

\item Manual testing of concurrency issues, by giving the tester an ability to control the scheduling as part of the test

\end{enumerate}

The first topic is the well studied one. The authors of different works focus on recording different types of events during the application execution and use the collected data in 
order to create the exact replay of recorded run. The examples of such works are \cite{carvertai}, \cite{pakam}, \cite{russinovich}, \cite{konuru}. The approach is so well studied that 
it is already utilized by commercial companies that provide production ready tools for record and replay of concurrent applications \cite{replaydirector}. Although this technique is very powerful for debugging and bug fixing purposes, it is less useful for testing since the bug prone run has to be somehow reproduced before it could be recorded for the first time.

The second group of the works is very heterogeneous. Many authors apply static analysis techniques to discover such types of concurrent bugs like deadlocks \cite{naik} or dataraces \cite{kahlon}. There is plenty of researches and tools \cite{compare1}, \cite{compare2}, \cite{compare3} implementing different analysis techniques. 

Other authors perform the analysis based on the data collected during the run. O'Callahan and Coi \cite{ocallahan} analyze the runtime
behavior of the application and apply lockset-based and happens-before techniques in order to identify potential bugs. Eraser \cite{eraser} 
tracks application actions and uses collected data to detect possible dataraces. RaceTrack \cite{racetrack} is another tool that utilizes
this approach but applies different algorithms in order to identify data races.

Another set of tools interfere with the threads scheduler work, forcing
the execution of uncommon executions flows. ConTest \cite{contest}
introduces new context switches into the program code thus revealing
hidden bugs. ConCrash \cite{concrash} utilizes record and replay
technique in order to reproduce buggy runs. AtomFuzzer \cite{atomfuzzer}
forces context switches inside critical regions trying to cause
atomicity violation. Microsoft Chess \cite{chess1, chess2} reruns each
test multiple times while enumerating over different possible thread
schedulings. 

DataCollider \cite{datacollider} is the only tool we are aware of that
makes use of the breakpoints mechanism. It breaks the execution on access
to random memory locations and analyzes the program state in order
to identify data races. Unlike \textit{Interleaving}, it does not use this mechanism in order to change the execution flow induced by OS threads scheduler.

All the techniques above are fully automated and do not make any use of the knowledge the developer has regarding his code.

The third group is the smallest one and contains only few researches. All the works in this group try to utilize some information provided by developer / tester in order to reproduce the buggy state.
ConAn \cite{conan1, conan2} and MultithreadedTC \cite{multithreadedtc}
split the application execution timeline to several slots providing
the developer the ability to order the code blocks with respect to
those slots. IMUnit \cite{imunit} introduces the concept of events
that occur during the test run and enforces events ordering specified
for the test. This technique is very close to the one we propose.
The comparison of our work to IMUnit was presented earlier in the
paper (section \ref{sec:imunitcompare}). Park and Sen \cite{park}
use the information provided by the developer regarding the buggy state
and try to enforce the scheduling that will reach this state.

\section{Conclusions}
\label{conclusions}

Testing concurrent applications is a very challenging task. One of the
reasons for this is lack of control over threads scheduling during
test execution and inability to reproduce the bug as the result
of this. We propose a novel technique that allows the unit test
developer to specify the desired threads scheduling as part of test
setup. This scheduling will be enforced during the test execution
consequently reproducing the bug on every test execution.

Our technique utilizes the breakpoints mechanism which allows us to preempt
the flow in arbitrary points in the code, including code under
test and third party libraries, without the need for code modification.
We also allow the test developer to define the decision logic for
every particular context switch using Java programming language. All
this makes our framework very powerful but still easy to learn and
use.

In order to demonstrate the feasibility of the technique we propose, we implemented a prototype of our ideas in the \textit{Interleaving} framework. Using this prototype we were able to reproduce several real life bugs that were considered hard to reproduce  till now. In addition, the framework has good integration with Eclipse IDE and JUnit and does not require dedicated runtime environment. Although the framework is implemented using Java, the technique itself is not bound to a specific language and can be implemented for other platforms too. 

We provide the comparison of our framework against the best similar tool we are familiar with. We show that \textit{Interleaving} has additional value when unit testing the application, allowing the tester to reproduce bugs that he could not reproduce using another tool. In addition we show that any unit test created for the other tool could be easily migrated to \textit{Interleaving} notation.


We believe our technique is promising and could be combined with
other works to achieve even better results. For example, the declarative
notation of IMUnit could be combined with the freedom that \textit{Interleaving}
provides to initiate the events from every place in the code. Moreover,
the idea of using breakpoints for execution flow interception could
be used for other purposes like invariants validation or code instrumentation.

\bibliographystyle{abbrv}

\end{document}